\documentclass[reprint,
amsmath,amssymb,aps,pra,
showpacs
]
{revtex4-1}

\usepackage{graphicx}
\usepackage{psfrag}
\usepackage{wrapfig}
\usepackage{amsfonts}  
\newcommand{\sech}[0]{{\rm sech}\; }
\usepackage{hyperref}

\begin{document}
\begin{abstract}
We present the results of asymptotic and numerical analysis of dissipative Kerr solitons in whispering gallery mode microresonators influenced by higher order dispersive terms leading to the appearance of a dispersive wave (Cherenkov radiation).  Combining direct perturbation method with the method of moments we find expressions for the frequency, strength, spectral width of the dispersive wave and soliton velocity. Mutual influence of the soliton and dispersive wave was studied. The formation of the dispersive wave leads to a shift of the soliton spectrum maximum from the pump frequency (spectral recoil), while the soliton displaces the dispersive wave spectral peak from the zero dispersion point.
\end{abstract}
\title{Dissipative Kerr solitons and Cherenkov radiation in optical microresonators with third order dispersion}

\author{A.V. Cherenkov$^{1,2}$}
\author{V.E. Lobanov$^1$}
\author{M.L. Gorodetsky$^{1,2}$}
\affiliation{$^1$Russian Quantum Center, Skolkovo 143025, Russia}
\affiliation{$^2$Faculty of Physics, Lomonosov Moscow State University, Moscow 119991, Russia }

\email[e-mail]{ mg@rqc.ru}
\maketitle

\section{Introduction }
The discovery of optical frequency combs in nonlinear whispering gallery mode (WGM) microresonators \cite{del2007optical} suggests a possibility of the development of novel types of frequency comb sources with characteristics (especially, compactness  and repetition rates) unachievable for the systems based on conventional mode-locked lasers \cite{28051}. This is of particular importance for various applications, such as precision frequency metrology \cite{PhysRevX.3.031003}, highly multiplexed spectroscopy \cite{diddams2007molecular}, low-noise microwave generation \cite{Quinlan:13}, terabit telecommunication \cite{Koos1}, laser ranging \cite{Koos2} and others \cite{kippenberg2011microresonator}.
By now such so-called Kerr frequency combs have been demonstrated in microresonators made of different materials, including  silica \cite{del2007optical}, fluoride crystals \cite{wang2013mid,lecaplain2016mid,Grudinin:16}, fused silica \cite{PhysRevX.3.031003}, silicon nitride  \cite{levy2010cmos}, diamond \cite{hausmann2014diamond}, aluminum nitride \cite{Jung:13}, just to name a few.

It was revealed, that the generation of frequency combs in nonlinear microresonators results from the cascaded four-wave mixing processes  \cite{PhysRevLett.93.083904,PhysRevLett.93.243905}. The line spacing of a microresonator-based frequency comb is determined by the microresonators free spectral range (FSR, the inverted round-trip time of light in the microresonator), which can range in GHz or THz domains. Nonlinear process of comb formation generally leads to the arbitrary phase relations between individual spectral lines, that is quite different from conventional laser-based frequency combs, and results in the emergence of significant phase noise of RF beatnote \cite{herr2012universal}. The generation of dissipative Kerr solitons (DKS) allows to solve this problem and opens a way to mode-locked coherent,  broadband optical frequency combs with smooth spectral profile  \cite{herr2014temporal}. Such solitons have been already demonstrated experimentally in optical crystalline \cite{herr2014temporal} and silica \cite{Yi:15} WGM microresonators as well as in integrated microrings \cite{brasch2016photonic}.

It was shown that the combined material and geometrical group velocity dispersion (GVD) of microresonators has strong influence on the possibility of soliton generation and its properties \cite{herr2014mode,yang2016broadband}. In particular it was predicted via numerical simulations \cite{coen2013modeling,Zhang:13,Wang:14,Hansson:14,milian2014soliton} that the process of the dispersive wave formation (optical analog of Cherenkov radiation) caused by the higher order dispersion terms \cite{Akhmediev:95} may expand the comb generation bandwidth into the normal dispersion regime \cite{brasch2016photonic}.
In \cite{brasch2016photonic} experimental generation of the soliton Kerr comb exploiting the dispersive wave and covering 2/3 of an octave was demonstrated for the first time in on-chip ring SiN microresonator. 

The formation of the dispersive wave may be explained as follows: if a microresonator is pumped at a laser pump wavelength characterized by the anomalous mocroresonator GVD, temporal solitons can be generated. If the duration of the soliton is short enough so that its bandwidth extends to the normal dispersion regime, near the point where the GVD is close to zero, the wavelength matching becomes especially favorable for four-wave mixing processes producing sharp spectral peak, corresponding in time domain to oscillating soliton tails -- dispersive waves \cite{coen2013modeling}. Smooth and broadened dispersive wave spectral peak can also be observed for incoherent Kerr combs \cite{Foster:11}. 

An interplay between cascaded four-wave mixing processes and dispersive wave formation was analysed in \cite{erkintalo2012cascaded}. Although phase-matching condition in  four-wave mixing allows the generation of a dispersive wave, this will not guarantee the formation of mode-locked soliton combs. Therefore, in spite of practical interest and  many attempts, mostly based on numerical simulations, the mutual influence of the soliton and dispersive wave, especially in the case of Kerr combs characterized by periodic boundary conditions, is not well understood. 

Dynamics of such process in space-time representation can be described by the Lugiato-Lefever equation (LLE) \cite{PhysRevLett.58.2209} with higher order dispersion terms \cite{milian2014soliton}. Soliton dynamics in presence of higher order dispersion \cite{PhysRevA.88.035802,Genty2010989} have been also studied  earlier with respect to supercontinuum generation \cite{PhysRevLett.101.113904,Dudley:09} and spectral broadening in optical fibers \cite{RevModPhys.82.1287}. The presence of the third order perturbation leads to the approximate solution in a form of a  soliton pulse with a radiation tail \cite{Akhmediev:95,Wang:14}. The actual spectral position of the dispersive wave peak, corresponding to the characteristic frequency of the tails' oscillations was analyzed in \cite{milian2014soliton}, and also considering Raman effect \cite{milian2015solitons}. Previous analysis was limited, however, as only the position of the dispersive wave was investigated. Hereby we perform the complete asymptotic  analysis of the system which determines the position, strength and spectral width of the dispersive wave as well as the spectral recoil produced by this wave.

The dynamics of the temporal dissipative cavity Kerr solitons in microresonators can be described using the damped driven nonlinear Schr\"odinger equation frequently referred as LLE  with higher order dispersion terms \cite{PhysRevA.87.053852,Santhanam2003413}. This equation is in fact an equation for the envelope of the optical field in a coordinate frame circulating in a microresonator with the pump field phase velocity:
\begin{equation}
i\frac{\partial\Psi}{\partial T}+\frac{1}{2}\frac{\partial^2 \Psi}{\partial z^2} +  \vert\Psi\vert^2\Psi-\zeta_{0}\Psi=-i \Psi+\sum_{k>2}(-i)^k d_k \frac{\partial^k \Psi}{\partial z^k}+if.
\label{lle}
\end{equation}
Here $\Psi$ is the slowly varying waveform, $\kappa$ is the energy damping coefficient. $T=\frac{\kappa t}{2}$ denotes the normalized time, $\eta$ is the coupling efficiency, $f=\sqrt{\frac{8 \eta g P_\mathrm{in}}{\kappa^2 \hbar\omega_0}}$ is the dimensionless pump amplitude for the pump power $P_\mathrm{in}$ with the nonlinear coupling coefficient $g=\frac{\hbar\omega_0^2 c n_2}{n_0^2 V_\mathrm{eff}}$, $n_0$ and $n_2$ are linear and nonlinear refractive indices of the material, and $V_\mathrm{eff}$ is the effective nonlinear mode volume. $\zeta_0=2\frac{\omega_0-\omega_p}{\kappa}$ is the normalized detuning from resonance. Higher order dispersion coefficients $d_k=D_k \frac{2}{\kappa k!}(\frac{\kappa}{2D_2})^{k/2}$, $k>2$ may be found from polynomial approximation for the ``cold'' cavity nearly equidistant eigenfrequencies of a mode family of interest $\omega_\mu=\omega_0+\sum_{k} \frac{D_k}{k!}\mu^k$, where  $\mu$ is the mode number defined in relation to the pumped mode  $\omega_0$.  Lower order coefficients are already imprinted in the basic equation and boundary conditions as $D_1 \simeq \dfrac{c}{n_0 R}$ is the FSR ($R$ is the radius of the resonator) and $D_2>0$ (anomalous dispersion) is eliminated from the second term using the substitution $z=\varphi\sqrt{\frac{\kappa}{2 D_2}}$, where $\varphi\in [0,2\pi]$ is the azimuthal angle. In experiment frequency comb is observed on optical spectrum analyzer as a sequence of equidistant lines with $|a_\mu|^2$
($a_\mu$ is the amplitude of the spectra line $\mu$), separated by FSR $D_1/2\pi$.

Without the right part equation \eqref{lle} is integrable with a known $\sech$--shaped soliton solution \cite{Zakharov&Shabat}.  
Though exact stationary solutions of equation \eqref{lle} with just a driving term but without losses and higher-order terms are also known \cite{Barashenkov&Smirnov}, this gives little insight in understanding of mutual interaction of the soliton and dispersive wave without extended numerical simulations and asymptotic approximations.

\section{Method of moments}
Bright dissipative solitons and corresponding coherent Kerr combs are possible only in far red detuned regime \cite{herr2014temporal} ($\zeta_0\gg1 $). Taking this fact into account, we introduce a small parameter $\gamma=\frac{1}{\zeta_0}$ and rewrite equation \eqref{lle} as follows:
\begin{equation}
i\psi_{\tau}+\psi_{xx}+2|\psi|^2\psi-\psi=-i\gamma\psi +\sum_{k>2} (-i)^k \delta_k\frac{\partial^k \psi}{\partial x^k} - \gamma H, 
\label{Blle}
\end{equation}
where $\tau=\zeta_0T$, $\psi=i\Psi/\sqrt{2\zeta_0}$, $x=z\sqrt{2\zeta_0}$, $H=f/\sqrt{2\zeta_0}$, $\delta_k = 2d_k (2\zeta_0)^{k/2-1}$. In this form equation \eqref{Blle} is very convenient for asymptotic analysis, though, as the important varying physical parameter detuning $\zeta_0$ is now incorporated in the scaling of both temporal and spatial coordinates, it is not very appealing  from the point of view of an experimentalist or for numerical simulations of the comb formation. That is why we are forced to jump several times in this paper between \eqref{Blle} and \eqref{lle}.

We use the method of moments to find the approximate soliton solutions of the LLE equation \cite{PhysRevE.73.036621}. The first five moments of the equation \eqref{Blle} are the following:
\begin{align}
{\cal M}_0&=\int\limits^\infty_{-\infty}|\psi|^2 \,dx = {\cal E},\label{integrals1}\\
{\cal D}_0&=\frac{i}{2}\int\limits^\infty_{-\infty}\left(\psi\psi^*_x-\psi^*\psi_x\right)\,dx = {\cal EK},\label{integrals2}\\
{\cal M}_1&=\int\limits^\infty_{-\infty}x|\psi|^2 \,dx = {\cal EX},\label{integrals3}\\
{\cal M}_2&=\int\limits^\infty_{-\infty}x^2|\psi|^2 \,dx= {\cal E}({\cal W}^2+{\cal X}^2),\label{integrals4}\\
{\cal D}_1&=\frac{i}{2}\int\limits^\infty_{-\infty}x\left(\psi\psi^*_x-\psi^*\psi_x\right)\,dx = {\cal E}({\cal KX}+{\cal W}^2{\cal C}).\label{integrals5}
\end{align}
The  integration limits are set as infinite because the soliton duration is much shorter than the soliton roundtrip time. These moments determine the pulse energy ${\cal E}$, the momentum ${\cal K}$, the position ${\cal X}$, the width ${\cal W}$ and the chirp ${\cal C}$ of a pulse. 
Taking time derivatives of the moments \eqref{integrals1}--\eqref{integrals5}, and substituting  time derivatives from (\eqref{Blle}), using some algebra, and assuming that $\psi(\pm\infty)=\psi_0$, $\psi_x(\pm\infty)=0$, we get:
\begin{align}
\frac{\partial{\cal M}_0}{\partial \tau} = &-2\gamma {\cal M}_0+ 2\gamma H \int\limits^{+\infty}_{-\infty}\Im(\psi)\,dx,\label{moments1}\\
\frac{\partial D_0}{\partial \tau}=&-2\gamma D_0,\label{moments2}
\end{align}
\begin{align}
\frac{\partial {\cal M}_1}{\partial \tau} =& -2\gamma {\cal M}_1+2{\cal D}_0+2\gamma H\!\int\limits_{-\infty}^{+\infty}\!x \Im(\psi)\, dx\nonumber\\
&+\sum\limits_{p=2} \int\limits_{-\infty}^{+\infty}(2p-1)\, \delta_{2p-1} \left|\frac{\partial^{p-1}\psi}{\partial x^{p-1}}\right|^2 dx \nonumber\\
&- i \sum\limits_{p=2} \int\limits_{-\infty}^{+\infty} p\,\delta_{2p}\left[\psi,\psi^*\right]_p dx,\label{moments3}
\end{align}
\begin{align}
\frac{\partial {\cal M}_2}{\partial \tau}=&-2\gamma {\cal M}_2+4{\cal D}_1  +2\gamma H\int\limits_{-\infty}^{+\infty}\!x^2 \Im(\psi)\, dx\nonumber\\
&+\sum\limits_{p=2}\int\limits_{-\infty}^{+\infty} 2x(2p-1)\, \delta_{2p-1} \!\left|\frac{\partial^{p-1}\psi}{\partial x^{p-1}}\right|^2 \,dx\nonumber\\
&- \sum\limits_{p=2}\int\limits_{-\infty}^{+\infty}  i p\, \delta_{2p}\left[\psi,\psi^*\right]_p dx,\label{moments4}\\
\frac{\partial {\cal D}_1}{\partial \tau} =&-2\gamma {\cal D}_1+\int\limits_{-\infty}^{+\infty}\!\left(x \frac{\partial}{\partial x}|\psi|^4 +2 |\psi_x|^2\right) dx \nonumber\\-& \gamma H\int\limits_{-\infty}^{+\infty}\! x \Re (\psi_x)\, dx
+ 2\sum\limits_{p=2}\int\limits_{-\infty}^{+\infty} p\delta_{2p}\left|\frac{\partial^{p}\psi}{\partial x^{p}}\right|^2 \,dx\nonumber\\
&-\sum\limits_{p=2}\int\limits_{-\infty}^{+\infty}\frac{i}{2}(2p-1)\delta_{2p-1}\,\left[\psi,\psi^*\right]_p\,dx.
\label{moments5}
\end{align}
where $[u,v]_p=\frac{\partial^{p} u}{\partial x^{p}}\frac{\partial^{p-1}v}{\partial x^{p-1}}-\frac{\partial^{p-1}u}{\partial x^{p-1}}\frac{\partial^{p}v}{\partial x^{p}}$, $\Re(\psi)$ and $\Im(\psi)$ are real and imaginary part of $\psi$.

The evolution equations \eqref{moments1}--\eqref{moments5} may be used to find approximate solutions of the LLE \eqref{Blle} assuming a trial function (Ansatz) taken from known solutions of the unperturbed equation and augmented with additional parameters.

One can make several general conclusions based on equations \eqref{moments1}--\eqref{moments5}. Taking into account that for small higher order dispersion values soliton solution has a practically symmetric bell-shaped profile sitting on an almost unmodulated pedestal, so that $\psi$ is nearly an even function of $x$, one gets that $[\psi,\psi^*],  p\geq 1$ is an odd function of $x$ and, consequently, $\int\limits_{-\infty}^{+\infty}\left[\psi,\psi^*\right]_p dx=0$, $\int\limits_{-\infty}^{+\infty}x\left[\psi,\psi^*\right]_pdx\neq0$, $\int\limits_{-\infty}^{+\infty}\left|\frac{\partial^{p}\psi}{\partial x^{p}}\right|^2 dx\neq0$, $\int\limits_{-\infty}^{+\infty}x\left|\frac{\partial^{p}\psi}{\partial x^{p}}\right|^2 dx=0$. In this way, the even-order dispersion terms ($\delta_{2p}$) only affect the relation between the width and the amplitude of the soliton and the phase chirp. Odd-order dispersion terms ($\delta_{2p-1}$) influence on the soliton dynamics (change of the soliton position -- center of mass with time) and on the emergence of non-stationary solutions.

\section{Dissipative soliton with third order dispersion perturbation}
Henceforth we consider only the first higher order term (third-order dispersion).
\begin{equation}
i\psi_{\tau}+\psi_{xx}+2|\psi|^2\psi-\psi=-i\gamma\psi +i \delta_3\psi_{xxx} - \gamma H. \label{Neweq}
\end{equation}
First, we study the equation without losses and pump (zero-th order $\gamma$):
\begin{equation}
i\psi_{\tau}+\psi_{xx}+2|\psi|^2\psi-\psi= i \delta_3\psi_{xxx}.
\label{D30}
\end{equation}
We search for the first order perturbation over $\delta_3$   as it was done in \cite{Akhmediev:95}:
\begin{align}
\psi&=\sech\! \tilde x\,\, e^{i\delta_3 g(\tilde x)},\nonumber\\
\tilde x &= x-\delta_3\nu \tau = x - \frac{d{\cal X}}{d\tau}\tau,\label{Chi}
\end{align}
where $\nu$ is an unknown parameter of soliton velocity which we would like to determine. This leads to an equation for $f(\tilde x)$:
\begin{equation}
\frac{d^2 g}{d\tilde x^2 
}-2 \frac{d g}{d\tilde x 
}\tanh \tilde x-(5-\nu)\tanh \tilde x+6\tanh^3\tilde x=0,
\label{feq}
\end{equation}
with the solution $g=\frac{1}{2}(\nu+1)\tilde x -\frac{3}{2}\tanh \tilde x$.
Thus the soliton solution of equation \eqref{D30}  is 
\begin{equation}
\psi=\sech\!\tilde x\,\, e^{i\delta_3((\nu+1)\tilde x-3\tanh \tilde x)/2}.
\label{trial}
\end{equation}

We choose a trial solution taking into account \eqref{trial} in the form proposed in \cite{herr2015dissipative}:
\begin{equation}
\psi=(c+S(x,\tau))e^{i \varphi_{0}},
\label{test}
\end{equation}
 where $c e^{i\varphi_{0}}=\psi_0$  is the cw background, and $S=\psi(\tilde x)e^{i\chi}$ is the soliton Ansatz, accounting for $\delta_3$. Here $\chi$ is an unknown phase of a soliton attractor which we need to find as well as a factor $\nu$, defining the soliton velocity.  

We substitute the test solution \eqref{test} in the equations \eqref{moments1}--\eqref{moments3} and require conservation of energy $\cal{E}$ and momentum $\cal{K}$ for the combs' stability \cite{Wang:14} and determine the center of mass position $\cal{X}$. We are not using below the equations \eqref{moments4}--\eqref{moments5} characterizing soliton width and chirp. At the first glance the integrals in the momentum method when applied to solitons on background will diverge, however, this is not the case, because we choose the background $\psi_0$ to obey the equation \eqref{Blle}.
\begin{align}
-\frac{1}{\gamma}\frac{\partial \mathcal {E}}{ \partial \tau}=&\int\limits_{-\infty}^{+\infty}2c(c-H\sin\varphi_{0}
) dx+2\int\limits_{-\infty}^{+\infty}c(S+S^{*}) dx +\nonumber\\& i H\int\limits_{-\infty}^{+\infty}(Se^{i\varphi_{0}}-S^{*}e^{-i\varphi_{0}}+2\vert S \vert^2)dx=0,\label{E1}\\
\frac{i}{\gamma}\frac{\partial \mathcal{K} }{\partial \tau}=&\frac{1}{\mathcal {E}}\int\limits_{-\infty}^{+\infty}(cS^{*}_x-cS_x+S S^{*}_x-S^{*}S_x)dx=0,\label{E2}
\end{align}
\begin{align}
\frac{\partial \mathcal {X}}{\partial \tau}=&\frac{1}{\mathcal {E}}\frac{\partial}{\partial \tau} \int\limits_{-\infty}^{+\infty}(x\vert S \vert^2 + cx(S^{*}+S))\,dx\label{E3}\\
=&-\frac{\gamma}{\mathcal {E}} \int\limits_{-\infty}^{+\infty}2xc(c- H \sin\varphi_{0} )dx\nonumber\\&-\frac{\gamma}{\mathcal {E}}\int\limits_{-\infty}^{+\infty}x(2\vert S \vert^2+iH(Se^{i\varphi_{0}}-S^{*}e^{-i\varphi_{0}})\,dx\nonumber\\
&- \frac{\gamma}{\mathcal {E}}\int\limits_{-\infty}^{+\infty} 2 c(S+S^{*}))dx + \frac{3\delta_{3}}{\mathcal {E}}\int\limits_{-\infty}^{+\infty}S_{x}^{*}S_{x}dx\nonumber.
\end{align}
In the above equations we select the terms which contain only cw component and equate them to zero: 
\begin{equation}
c - H\sin\varphi_{0}=0. 
\label{zerint}
\end{equation}
This requirement allows avoiding divergence.
The constant  $c$  and the phase $\varphi_0$  are  determined by the solution of the algebraic equation  obtained by substituting cw $\psi_0$ in \eqref{Blle}: 
\begin{align}
(2c^2-1+i\gamma)ce^{i\varphi_0}=-\gamma H. 
\label{cweq}
\end{align}
By moving the complex exponent in \eqref{cweq} to the right and looking at imaginary parts we confirm that \eqref{zerint} is  satisfied automatically.
For the large detuning  $\gamma\ll1$ and $\vert\psi_0\vert^2\ll1$ \cite{herr2015dissipative} we get
\begin{align}
c&=\gamma H+\frac{1}{2}H(4H^2-1)\gamma^3+O(\gamma^5),\nonumber\\
\sin\varphi_0&=\gamma+\frac{1}{2}(4H^2-1)\gamma^3+O(\gamma^5).
\end{align}
From \eqref{E1} we obtain:
\begin{equation}
  \sin(\chi+\varphi_0)- 2\gamma\cos(\chi)=\frac{2}{\pi H},
  \label{P1}
\end{equation}
then finally from \eqref{E3}:
\begin{equation}
\nu(1+\pi c \cos\chi)=1.
\label{P2}
\end{equation}
The equation \eqref{P1} provides the soliton phase in agreement with earlier analysis \cite{Nozaki1984,Wabnitz1993}:
\begin{equation}
\chi=\arcsin\frac{2}{\pi H\sqrt{1+\gamma^2}  }+\gamma\approx \arcsin(\frac{2}{\pi H})+\gamma+O(\gamma^2).
\label{ph}
 \end{equation}
 The soliton phase is therefore determined by the pump detuning and power.

From the equations  \eqref{P2} in view of \eqref{ph} and  \eqref{E2} we obtain:
\begin{equation}
\nu\approx 1 +O(\gamma +d_3). \label{nuone}
\end{equation}
In this way, the velocity of the center of mass of the soliton in current coordinates is equal to $\delta_3$.
The soliton on background approximate solution in a more physical form of \eqref{lle} is therefore:  
\begin{equation}
\Psi=\sqrt{2\zeta_0}\,\sech\!(\sqrt{2\zeta_0}\tilde{z})\,e^{i d_3 g(z)+
i\arccos({\sqrt{8\zeta_0}}/{\pi f })}-i\frac{f}{\zeta_0},
\label{solution}
\end{equation}
where $\tilde{z}=z-2d_3\zeta_0T$ and  $g(z)=(4\zeta_0 \tilde{z}-
3\sqrt{2\zeta_0}\tanh(\sqrt{2\zeta_0}\tilde{z}))$.
In this way the third-order dispersion term leads to the  soliton motion with the constant velocity  dependent on the detuning and the third-order dispersion value:
\begin{equation}
\nu_{z}=2d_3 \zeta_0.
\label{velocity1}
\end{equation}
Nonzero soliton velocity has an important consequence. Since the ``transverse'' coordinate $\varphi$ (or $z$ in terms of Eq. \eqref{lle}) is defined in coordinate system rotating with angular velocity $D_1$ ($\varphi=\phi-D_1t$, where $\phi$ is the regular polar angle inside the microresonator), the soliton velocity associated with the third order dispersion may be interpreted as the shift of the soliton repetition rate (in the absence of the third order dispersion, the soliton repetition rate is equal to $D_1/2\pi$). Consequently, using  Eq. \eqref{velocity1} and taking normalization procedure into account one can get the expression for the soliton repetition rate shift: $2\pi \Delta f_r=\frac{d\varphi}{dt}=\zeta_0\frac{\kappa D_3}{6D_2}$. In this way the repetition rate may be tuned by laser detuning which may be used in practical application of soliton comb based microwave oscillators.

\begin{figure}[h!]
\center{\includegraphics[trim=10 20 50 0,clip,width=1\linewidth]{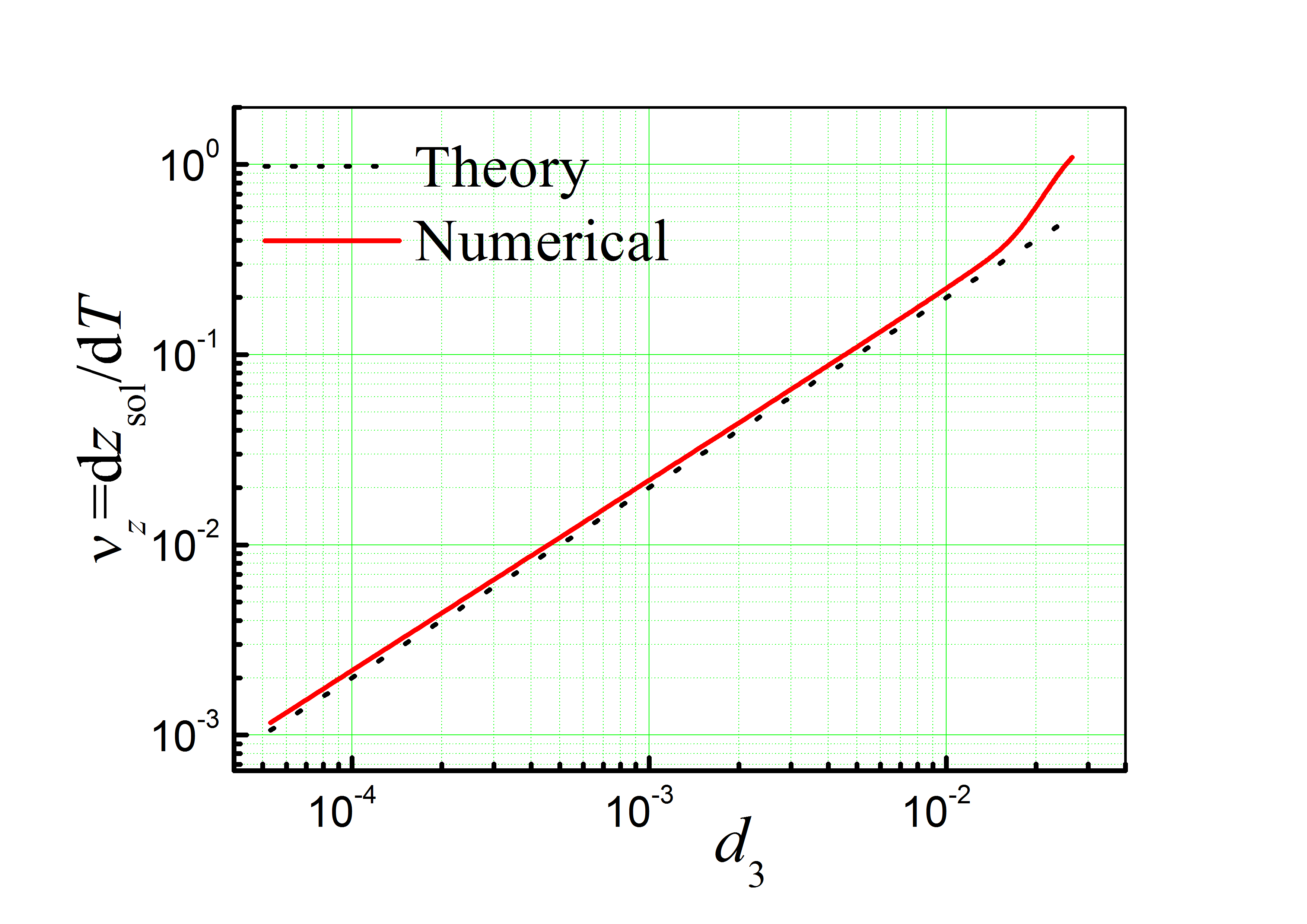} }
\center{\includegraphics[trim=10 20 50 0,clip,width=1\linewidth]{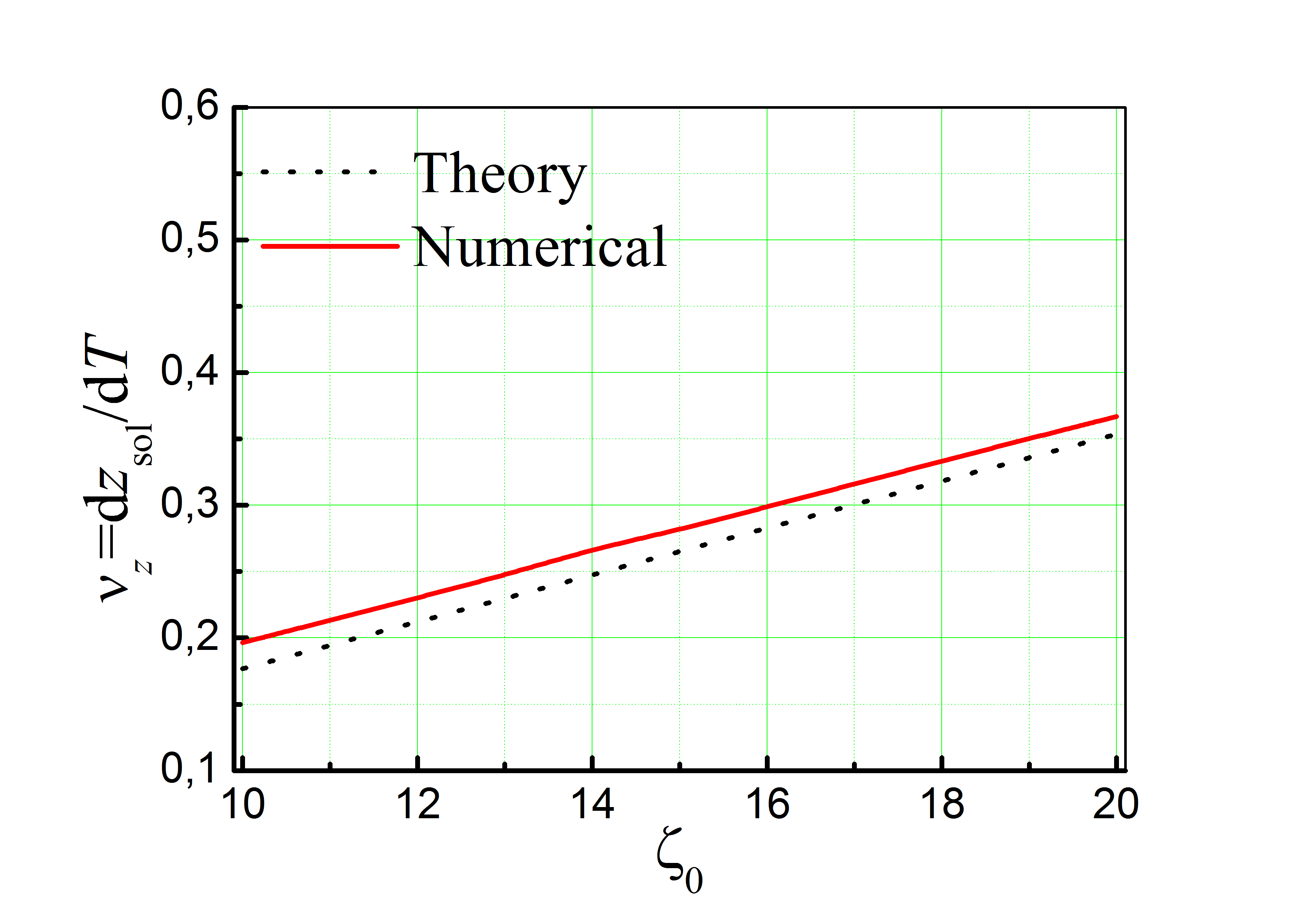} }
\caption{Comparison of analytical \eqref{velocity1} (dotted line) and numerical (solid line) results. (Top panel): soliton velocity versus  third-order dispersion parameter $d_3$ for $\zeta_0=10$, and soliton velocity versus detuning $\zeta_0$  for $d_3=0.0088$ ($D_3/D_2=0.0075$) (bottom). In all cases $f=4.1$, $D_2/\kappa=0.01$.}
\label{1}
\end{figure}
As one can see from Fig. \eqref{1}, the obtained  approximate solution is in a good agreement with the numerical simulation results  in a wide range of experimentally possible parameters. Note, that the obtained  result may be further refined if  higher order corrections (nonlinear terms of $\gamma$) are taken into account in evolution equations \eqref{E1}, \eqref{E2}, \eqref{E3}.

To find an approximate perturbed soliton spectrum we expand the phase with $\tanh$--term of \eqref{solution} into a series over third-order dispersion term, leaving the first term in expansion and use Fourier transform:
\begin{align}
\mathcal{F}\left[(\sqrt{2\zeta_0}\,\sech\!(\sqrt{2\zeta_0}\tilde z)\,e^{ i 4d_3\zeta_0 \tilde z}\right.\nonumber\\ \left.\times(1-i3d_3\sqrt{2\zeta_0}\tanh(\sqrt{2\zeta_0}\tilde z))\right].
\end{align} 
Taking into account that $z=\varphi\sqrt{\frac{1}{2d_2}}$ and
\begin{align}
{\mathcal F}[S]=\int S(\varphi)e^{-i\mu\varphi} d\varphi,
\end{align} 
 we get
\begin{align}
\mathcal{F}[S]=&e^{-id_3\zeta_0\sqrt{8d_2}T\mu}(\mathcal{F}S_0(\mu-\mu_r)\nonumber\\&-d_3\sqrt{18d_2}(\mu-\mu_r)\mathcal{F}[S_0(\mu-\mu_r))],
\label{spectr}
\end{align}
where $\mu_r=\frac{4d_3\zeta_0}{\sqrt{2d_2}}$, $\mathcal{F}S_0=\sqrt{d_2/2}\,\sech(\frac{\pi \mu}{2}\sqrt{\frac{d_2}{\zeta_0}})$ is unperturbed soliton solution spectrum.

It is clear from the Eq. \eqref{spectr} that the third order dispersion leads to the shift of the  soliton  maximum to higher frequencies. Taking into account that $d_2=\frac{D_2}{\kappa}$, $d_3=\frac{D_3\sqrt{\kappa}}{3\sqrt{8D_2^3}}$ one may estimate the position of the maximum of soliton spectrum $\mu_r$ (soliton recoil) as
\begin{equation}
\mu_r=\zeta_0\frac{D_3 \kappa}{3D_2^2}.
\label{maxD}
\end{equation}
The intensity of the spectral component at the point of the soliton maximum can be obtained from \eqref{spectr}
\begin{equation}
|a_{\mu_r}|^2=|\mathcal{F} \psi(\mu_r)|^2\approx \frac{D_2}{2\kappa}.
\end{equation}
Therefore, the intensity of the soliton maximum does not depend on third order dispersion parameters in linear approximation.

The proposed method  works well for large detuning enough ($\zeta_0>>1$) and low cw background ($f/\zeta_0<<1$). However, the maximexistenceal detuning providing soliton existence is $\zeta_{0max}=\frac{\pi^2 f^2}{8}$ \cite{herr2015dissipative}. Therefore, if pump power decreases, the maximal detuning value also decreases. So, if the ratio of the soliton peak intensity to cw  background intensity $(\frac{2\zeta^3_0}{f^3})$ \cite{herr2014temporal} is relatively small, we can not neglect the nonlinear interaction of the soliton with the background.
\section{Dispersive Wave}
\begin{figure}[h!]
\center{\includegraphics[trim=0 340 0 180,clip,width=\linewidth]{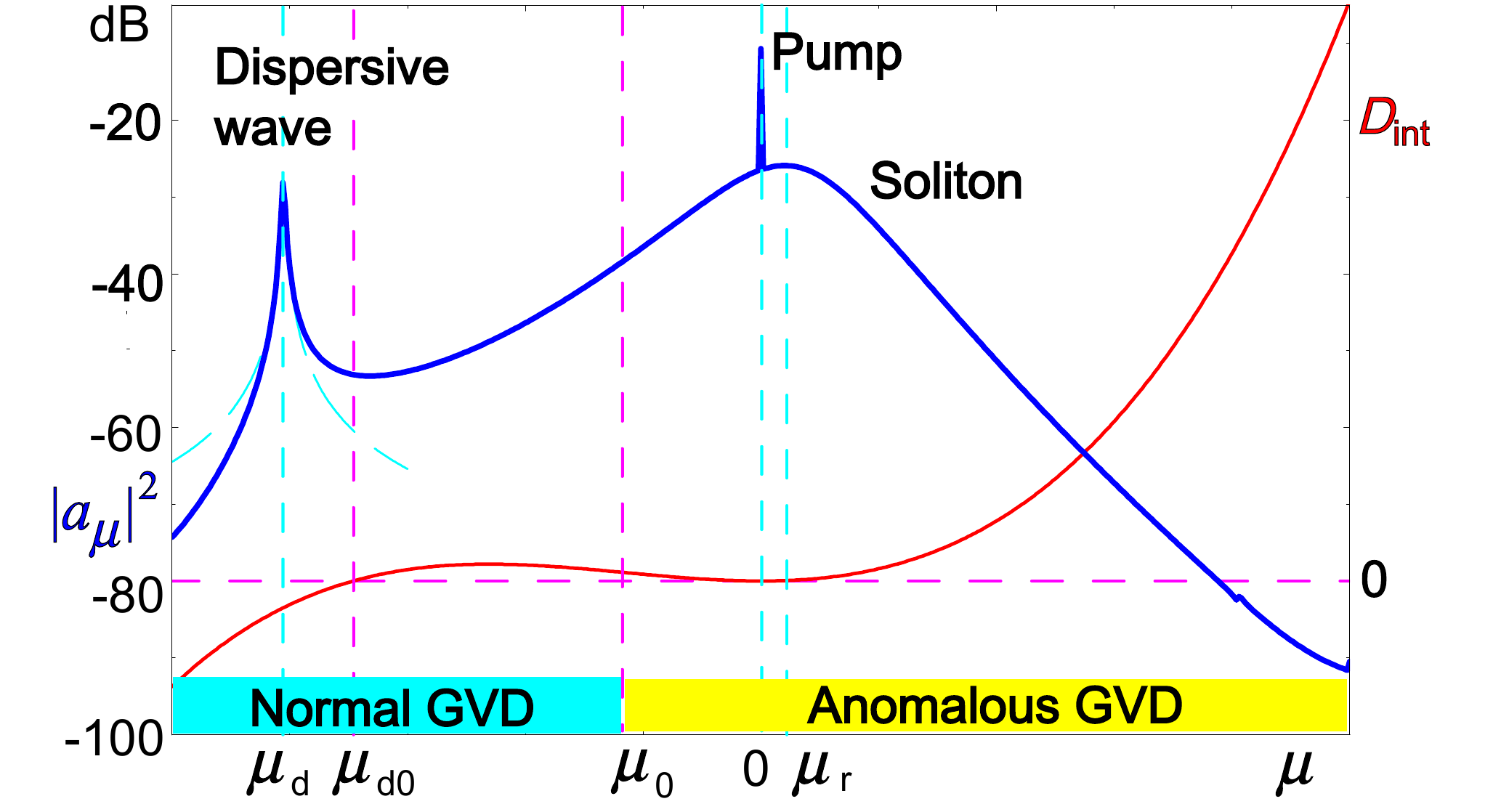} }
\caption{Schematic image of the  dispersion law and associated  soliton and dispersive wave   dynamics. $D_\mathrm{int}$ describes integrated dispersion and $D_\mathrm{int}(\mu_{d0})=D_\mathrm{int}(0)=0$, $D_\mathrm{int}^{''}(\mu_0)=0$. $\mu_{d0}$ is zero dispersion point, $\mu_r$ is soliton  spectral maximum position, $\mu_d$ is dispersive wave position. Pump position corresponds to $\mu=0$.}
\end{figure}

Third-order GVD leads to the emission of the resonant radiation by solitons that may be interpreted as the emergence of Cherenkov radiation \cite{Akhmediev:95}. In this part we analyze the parameters of the dispersive wave neglecting dispersive terms higher than the third order. To account for the difference of stationary soliton group velocity caused by the higher order dispersion terms we make the following change of variables in equation \eqref{Neweq}, taking into account \eqref{Chi},\eqref{nuone}:
\begin{equation}
\tilde{x}=x-\delta_3\tau.
\end{equation}

The initial stationary equation \eqref{lle} in these variables takes the following form:
 \begin{equation}
-i\delta_3\frac{\partial\psi}{\partial \tilde{x}}+\frac{\partial^2 \psi}{\partial\tilde{x}^2} +  2\vert\psi\vert^2\psi-\psi=-i \gamma\psi+i \delta_{3}\frac{\partial^3 \psi}{\partial\tilde{x}^3}+\gamma H.
\label{linequ}
\end{equation}
To search for the dispersive  wave parameters  we represent the dispersive wave as a harmonic solution with background \cite{Afanasjev:96,RevModPhys.82.1287}:
\begin{equation}
\psi=c+g_{1}e^{-iq\tilde{x}}+g_{2}^*e^{iq^*\tilde{x}}.
\end{equation}
We linearize the equation \eqref{linequ} for the  small amplitudes $g_{1,2}$ and get the system of two equations:
\begin{align}
 \begin{cases}
(\delta_3(q^3-q)-q^2-1+i\gamma+4|c|^2)g_{1}+2c^2g_{2}=0,
\\
2c^{*2}g_{1}-(\delta_3(q^3-q)-q^2-1-i\gamma+4|c|^2)g_{2}=0.
\end{cases}
\label{s1}
\end{align}
The condition of solvability of the linear system \eqref{s1} can be written as:
\begin{align}
(4|c|^2-q^2-1+i\gamma+\delta_3(q^3-q))\nonumber\\ \times(4|c|^2-1-i\gamma-q^2-\delta_3(q^3-q))=4|c|^4.
\end{align}
Assuming that the soliton background is weak for strongly red detuned case ($\gamma\ll1$) when solitons are possible, $c\simeq \gamma H$, $\delta_3\ll1$ and wavenumber $q$ is large, we find the zero-order approximation $q_0=\pm{\delta_3^{-1}}$.  Two dispersive resonances are symmetric with respect to the pump frequency. The second resonance may be interpreted as a result of parametric interaction between the pump and radiation peak (obeying $2\hbar\omega_{pump}=\hbar\omega_{d}+\hbar\omega_{-d}$). However, the amplitude of the resonance, corresponding to positive $q^{(0)}$ is relatively small, because phase-matching conditions are not satisfied due to strong dispersion in this point. Finding the next perturbation order $q\simeq q^{(0)}+\delta_3 q^{(1)}$, we'll get:
\begin{equation}
q= -\frac{1}{\delta_3}-\delta_3(2-4\gamma^2 H^2-i\gamma). 
\end{equation}
If we now turn to angular numbers $q\tilde x = \mu\tilde\phi$, we obtain the spectral position of the dispersive wave peak in terms of wavenumber $\mu$ (or frequency $\mu D_1/2\pi$): 
\begin{equation}
\mu_d=-\frac{3D_2}{D_3}-\frac{D_3\kappa}{3D_2^2}(2\zeta_0-2(f/
\zeta_0)^2-i).
\label{mu}
\end{equation}
Note that  the position of the dispersive wave is shifted from the zero dispersion point $\mu_{d0}=-\frac{3D_{2}}{D_{3}}$, corresponding to a zero value of the integrated dispersion $D_{int}=\omega_\mu-\omega_0-D_1\mu$ \cite{brasch2016photonic}:
\begin{equation}
\delta\mu_d=\Re(\mu_d)-\mu_{d0}\approx-\frac{2 D_3\kappa}{3D_2^2}\zeta_0.
\label{recoil1}
\end{equation}
In this way, the formation of the dispersive wave leads to a shift of the soliton spectrum maximum from the pump frequency \eqref{maxD}, as well as the soliton leads to the shift of the dispersive wave spectral position from the zero dispersion point \eqref{recoil1}. This phenomenon can be interpreted as the emergence of the spectral recoil. Note, that in order to find the shifted position of the dispersive wave we took into account the soliton velocity appeared due to the third order dispersion found using the soliton momentum method. 

\begin{figure}[h!]
\center{\includegraphics[trim=10 5 50 0,clip,width=1\linewidth]{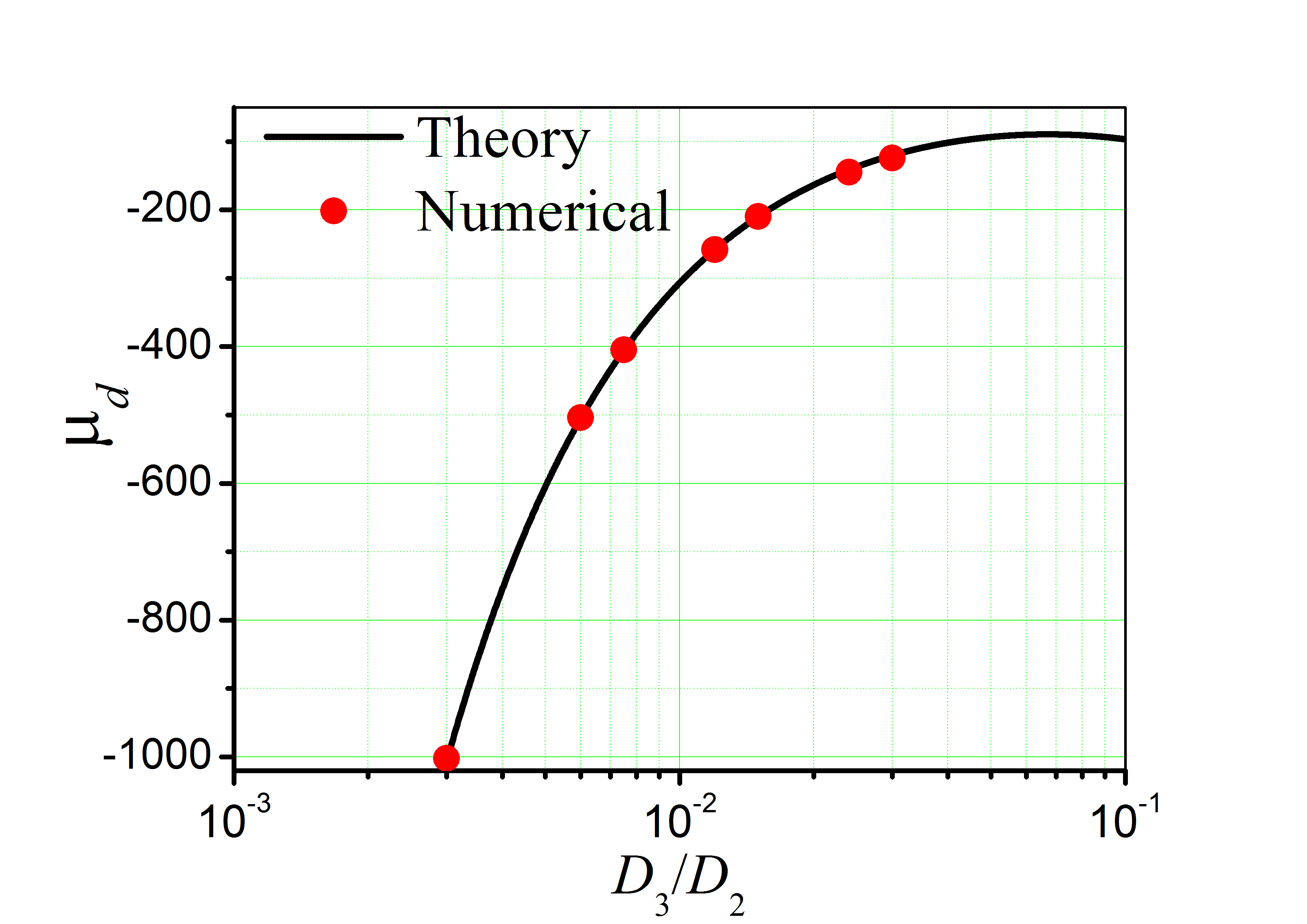} }
 \center{\includegraphics[trim=10 5 50 0,clip,width=1\linewidth]{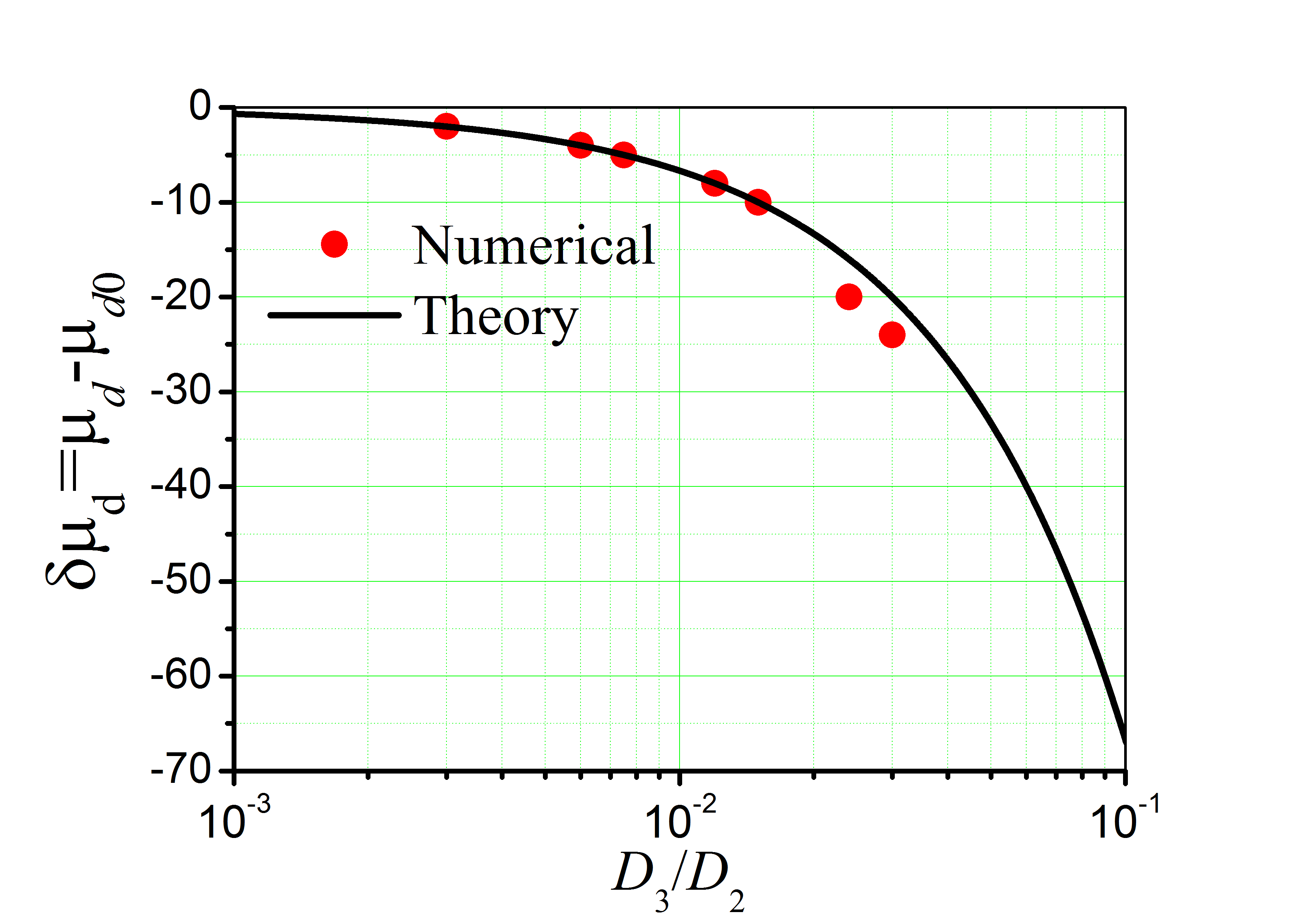} }
\caption{Comparison of analytical and numerical results. (Top): dispersive wave spectral maximum position vs.  dispersion parameters ratio $D_3/D_2$, and (bottom): dispersive wave recoil  vs.  dispersion parameters ratio $D_3/D_2$. In all cases $\zeta_0=10$, $f=4.1$, $D_2/\kappa=0.01$.}
\label{2}
\end{figure}

We compared the results of numerical solution of \eqref{lle} and theoretical predictions \eqref{mu},\eqref{recoil1} and found that they are in good agreement (see Fig.\eqref{2}). 
Note, that this approximation for the dispersive wave recoil works well if the following condition is satisfied:
\begin{equation}
\frac{D_3^2\kappa}{D_2^3}\zeta_0\ll 1.
\label{condition1}
\end{equation}
The imaginary part of \eqref{mu} describes the width of the Lorentzian  radiation peak:
\begin{equation}
\Delta_d=2\Im(\mu_d)=\frac{2D_3 \kappa}{3 D^2_2}.
\end{equation}
The dispersive wave in a microresonator may behave quite differently from the case of a fiber if $2\pi\times \frac{D_3 \kappa}{3 D^2_2}\sim 1$. In this case its tail may interfere constructively or destructively with its head and a standing sinusoidal pattern appears. This may lead to the new unexplored  effects and instabilities produced by the interaction between the soliton and the dispersive wave. This type of instability is observed in numerical simulations and may limit the possibility of a coherent comb expansion into normal dispersion region using dispersive wave emission.

The intensity of the spectral component at the dispersive wave maximum $\mu_d$ may be found  based on the conservation of the spectral center of mass \cite{herr2015dissipative}:
\begin{equation}
\sum_{\mu}\mu |a_{\mu}|^2=0.
\end{equation}
Dividing this sum into two parts, we obtain
\begin{equation}
\Re(\mu_d)\sum_\mu |a^{(d)}_\mu|^2+\mu_r\sum_{\mu} |a^{(s)}_\mu|^2\simeq 0
\label{amp}.
\end{equation}
From the equation  \eqref{amp} in view of the Lorentz shape of the dispersive wave spectrum we obtain:
\begin{align}
\Re(\mu_d) |a_d|^2\int\frac{1}{1+(\mu/\Im(\mu))^2}d\mu + \mu_r\int |\mathcal{F}S_0(\mu)|^2d\mu \simeq 0
\label{dmax}
\end{align}
Using \eqref{dmax} and taking \eqref{spectr} into account, we derive an expression for the dispersive wave peak spectral intensity:
\begin{equation}
|a_d|^2 \approx - \frac{2\mu_r \sqrt{D_2\zeta_0}}{\Im(\mu_d)\Re(\mu_d)\pi^2 \sqrt{\kappa}}=\frac{2D_3}{3\pi^2 \sqrt{\kappa D_2} } \zeta_0^{3/2}.
\end{equation}

In this way, the intensity at the dispersive wave maximum depends on detuning and third order dispersion parameters.  This fact may be used to control the magnitude of dispersive wave maximum.
 
\section{Conclusion}
	We performed a complete closed form analytical asymptotic analysis of dissipative Kerr solitons in microresonators with third dispersion. Perturbed soliton parameters were determined using the method of moments. It has been demonstrated that  the dispersive wave formation in the presence of a soliton contributes to expansion of the bandwidth of the generated comb to the normal dispersion frequency range. \textbf{A method of soliton repetition rate tuning was proposed.} 
\bigskip

\section{ACKNOWLEDGMENTS}
This work was supported by the Ministry of Education and Science of the Russian Federation (project RFMEFI58516X0005).

\bibliographystyle{apsrev4-1} 
\bibliographystyle{unsrt}
\bibliography{citations}

\begin{thebibliography}{10}

\bibitem{del2007optical}
P.~Del’Haye, A.~Schliesser, O.~Arcizet, T.~Wilken, R.~Holzwarth, and T.~J.
  Kippenberg.
\newblock Optical frequency comb generation from a monolithic microresonator.
\newblock {\em Nature}, 450(7173):1214--1217, 2007.

\bibitem{28051}
J.~Ye and S.~T. Cundiff, editors.
\newblock {\em Femtosecond Optical Frequency Comb: Principle, Operation, and
  Applications}.
\newblock Kluwer Academic Publishers, Boston, 2005.

\bibitem{PhysRevX.3.031003}
S.~B. Papp, P.~Del'Haye, and S.~A. Diddams.
\newblock Mechanical control of a microrod-resonator optical frequency comb.
\newblock {\em Phys. Rev. X}, 3:031003, Jul 2013.

\bibitem{diddams2007molecular}
S.~A. Diddams, L.~Hollberg, and V.~Mbele.
\newblock Molecular fingerprinting with the resolved modes of a femtosecond
  laser frequency comb.
\newblock {\em Nature}, 445(7128):627--630, 2007.

\bibitem{Quinlan:13}
F.~Quinlan, T.~M. Fortier, H.~Jiang, and S.~A. Diddams.
\newblock Analysis of shot noise in the detection of ultrashort optical pulse
  trains.
\newblock {\em J. Opt. Soc. Am. B}, 30(6):1775--1785, Jun 2013.

\bibitem{Koos1}
J.~Pfeifle, V.~Brasch, M.~Lauermann, Y.~Yu, D.~Wegner, T.~Herr, K.~Hartinger,
  P.~Schindler, J.~Li, D.~Hillerkuss, R.~Schmogrow, C.~Weimann, R.~Holzwarth,
  W.~Freude, J.~Leuthold, T.~J. Kippenberg, and C.~Koos.
\newblock Coherent terabit communications with microresonator {K}err frequency
  combs.
\newblock {\em Nat. Photon.}, 8(5):375--380, 2014.

\bibitem{Koos2}
C.~Weimann, M.~Lauermann, T.~Fehrenbach, R.~Palmer, F.~Hoeller, W.~Freude, and
  C.~Koos.
\newblock Silicon photonic integrated circuit for fast distance measurement
  with frequency combs.
\newblock {\em CLEO: 2014}, 2014.

\bibitem{kippenberg2011microresonator}
T.~J. Kippenberg, R.~Holzwarth, and S.~A. Diddams.
\newblock Microresonator-based optical frequency combs.
\newblock {\em Science}, 332(6029):555--559, 2011.

\bibitem{wang2013mid}
C.~Y. Wang, T.~Herr, P.~Del’Haye, A.~Schliesser, J.~Hofer, R.~Holzwarth,
  T.~W. H{\"a}nsch, N.~Picqu{\'e}, and T.~J. Kippenberg.
\newblock Mid-infrared optical frequency combs at 2.5 $\mu$m based on
  crystalline microresonators.
\newblock {\em Nat. Comm.}, 4:1345, 2013.

\bibitem{lecaplain2016mid}
C.~Lecaplain, C.~Javerzac-Galy, M.~L. Gorodetsky, and T.~J. Kippenberg.
\newblock Mid-infrared ultra-high-{Q} resonators based on fluoride crystalline
  materials.
\newblock {\em arXiv:1603.07305}, 2016.

\bibitem{Grudinin:16}
I.~S. Grudinin, K.~Mansour, and N.~Yu.
\newblock Properties of fluoride microresonators for mid-{IR} applications.
\newblock {\em Opt. Lett.}, 41(10):2378--2381, May 2016.

\bibitem{levy2010cmos}
J.~S. Levy, A.~Gondarenko, M.~A. Foster, A.~C. Turner-Foster, A.~L. Gaeta, and
  M.~Lipson.
\newblock {CMOS}-compatible multiple-wavelength oscillator for on-chip optical
  interconnects.
\newblock {\em Nat. Photon.}, 4(1):37--40, 2010.

\bibitem{hausmann2014diamond}
B.~J.~M. Hausmann, I.~Bulu, V.~Venkataraman, P.~Deotare, and M.~Lon{\v{c}}ar.
\newblock Diamond nonlinear photonics.
\newblock {\em Nat. Photon.}, 8(5):369--374, 2014.

\bibitem{Jung:13}
H.~Jung, C.~Xiong, K.~Y. Fong, X.~Zhang, and H.~X. Tang.
\newblock Optical frequency comb generation from aluminum nitride microring
  resonator.
\newblock {\em Opt. Lett.}, 38(15):2810--2813, Aug 2013.

\bibitem{PhysRevLett.93.083904}
T.~J. Kippenberg, S.~M. Spillane, and K.~J. Vahala.
\newblock {K}err-nonlinearity optical parametric oscillation in an
  ultrahigh-$q$ toroid microcavity.
\newblock {\em Phys. Rev. Lett.}, 93:083904, Aug 2004.

\bibitem{PhysRevLett.93.243905}
A.~A. Savchenkov, A.~B. Matsko, D.~Strekalov, M.~Mohageg, V.~S. Ilchenko, and
  L.~Maleki.
\newblock Low threshold optical oscillations in a whispering gallery mode
  {C}a{F}$_2$ resonator.
\newblock {\em Phys. Rev. Lett.}, 93:243905, Dec 2004.

\bibitem{herr2012universal}
T.~Herr, K.~Hartinger, J.~Riemensberger, C.~Y. Wang, E.~Gavartin, R.~Holzwarth,
  M.~L. Gorodetsky, and T.~J. Kippenberg.
\newblock Universal formation dynamics and noise of {K}err-frequency combs in
  microresonators.
\newblock {\em Nat. Photon.}, 6(7):480--487, 2012.

\bibitem{herr2014temporal}
T.~Herr, V.~Brasch, J.~D. Jost, C.~Y. Wang, N.~M. Kondratiev, M.~L. Gorodetsky,
  and T.~J. Kippenberg.
\newblock Temporal solitons in optical microresonators.
\newblock {\em Nat. Photon.}, 8(2):145--152, 2014.

\bibitem{Yi:15}
X.~Yi, Q.-F. Yang, K.~Y. Yang, M.-G. Suh, and K.~Vahala.
\newblock Soliton frequency comb at microwave rates in a high-{Q} silica
  microresonator.
\newblock {\em Optica}, 2(12):1078--1085, Dec 2015.

\bibitem{brasch2016photonic}
V.~Brasch, M.~Geiselmann, T.~Herr, G.~Lihachev, M.~H.~P. Pfeiffer, M.~L.
  Gorodetsky, and T.~J. Kippenberg.
\newblock Photonic chip--based optical frequency comb using soliton {C}herenkov
  radiation.
\newblock {\em Science}, 351(6271):357--360, 2016.

\bibitem{herr2014mode}
T.~Herr, V.~Brasch, J.~D. Jost, I.~Mirgorodskiy, G.~Lihachev, M.~L. Gorodetsky,
  and T.~J. Kippenberg.
\newblock Mode spectrum and temporal soliton formation in optical
  microresonators.
\newblock {\em Phys. Rev. Lett.}, 113(12):123901, 2014.

\bibitem{yang2016broadband}
Ki~Youl Yang, Katja Beha, Daniel~C. Cole, Xu~Yi, Pascal Del'Haye, Hansuek Lee,
  Jiang Li, Dong~Yoon Oh, Scott~A Diddams, Scott~B Papp, et~al.
\newblock Broadband dispersion-engineered microresonator on a chip.
\newblock {\em Nature Photonics}, 10:316--320, May 2016.

\bibitem{coen2013modeling}
S.~Coen, H.~G. Randle, T.~Sylvestre, and M.~Erkintalo.
\newblock Modeling of octave-spanning {K}err frequency combs using a
  generalized mean-field {L}ugiato--{L}efever model.
\newblock {\em Opt. Lett.}, 38(1):37--39, 2013.

\bibitem{Zhang:13}
L.~Zhang, C.~Bao, V.~Singh, J.~Mu, C.~Yang, A.~M. Agarwal, L.~C. Kimerling, and
  J.~Michel.
\newblock Generation of two-cycle pulses and octave-spanning frequency combs in
  a dispersion-flattened micro-resonator.
\newblock {\em Opt. Lett.}, 38(23):5122--5125, Dec 2013.

\bibitem{Wang:14}
S.~Wang, H.~Guo, X.~Bai, and X.~Zeng.
\newblock Broadband {K}err frequency combs and intracavity soliton dynamics
  influenced by high-order cavity dispersion.
\newblock {\em Opt. Lett.}, 39(10):2880--2883, May 2014.

\bibitem{Hansson:14}
T.~Hansson, D.~Modotto, and S.~Wabnitz.
\newblock On the numerical simulation of {K}err frequency combs using coupled
  mode equations.
\newblock {\em Opt. Commun.}, 312:134--136, 2014.

\bibitem{milian2014soliton}
C.~Mili{\'a}n and D.~V. Skryabin.
\newblock Soliton families and resonant radiation in a micro-ring resonator
  near zero group-velocity dispersion.
\newblock {\em Opt. Exp.}, 22(3):3732--3739, 2014.

\bibitem{Akhmediev:95}
N.~Akhmediev and M.~Karlsson.
\newblock {C}herenkov radiation emitted by solitons in optical fibers.
\newblock {\em Phys. Rev. A}, 51:2602--2607, Mar 1995.

\bibitem{Foster:11}
M.~A. Foster, J.~S. Levy, O.~Kuzucu, K.~Saha, M.~Lipson, and A.~L. Gaeta.
\newblock Silicon-based monolithic optical frequency comb source.
\newblock {\em Opt. Exp.}, 19(15):14233--14239, Jul 2011.

\bibitem{erkintalo2012cascaded}
M.~Erkintalo, Y.~Q. Xu, S.~G. Murdoch, J.~M. Dudley, and G.~Genty.
\newblock Cascaded phase matching and nonlinear symmetry breaking in fiber
  frequency combs.
\newblock {\em Phys. Rev. Lett.}, 109(22):223904, 2012.

\bibitem{PhysRevLett.58.2209}
L.~A. Lugiato and R.~Lefever.
\newblock Spatial dissipative structures in passive optical systems.
\newblock {\em Phys. Rev. Lett.}, 58:2209--2211, May 1987.

\bibitem{PhysRevA.88.035802}
M.~Tlidi, L.~Bahloul, L.~Cherbi, A.~Hariz, and S.~Coulibaly.
\newblock Drift of dark cavity solitons in a photonic-crystal fiber resonator.
\newblock {\em Phys. Rev. A}, 88:035802, Sep 2013.

\bibitem{Genty2010989}
G.~Genty, C.M. de~Sterke, O.~Bang, F.~Dias, N.~Akhmediev, and J.M. Dudley.
\newblock Collisions and turbulence in optical rogue wave formation.
\newblock {\em Physics Letters A}, 374(7):989 -- 996, 2010.

\bibitem{PhysRevLett.101.113904}
A.~Mussot, E.~Louvergneaux, N.~Akhmediev, F.~Reynaud, L.~Delage, and M.~Taki.
\newblock Optical fiber systems are convectively unstable.
\newblock {\em Phys. Rev. Lett.}, 101:113904, Sep 2008.

\bibitem{Dudley:09}
J.~M. Dudley, G.~Genty, F.~Dias, B.~Kibler, and N.~Akhmediev.
\newblock Modulation instability, akhmediev breathers and continuous wave
  supercontinuum generation.
\newblock {\em Opt. Express}, 17(24):21497--21508, Nov 2009.

\bibitem{RevModPhys.82.1287}
Dmitry~V. Skryabin and Andrey~V. Gorbach.
\newblock \textit{Colloquium} : Looking at a soliton through the prism of
  optical supercontinuum.
\newblock {\em Rev. Mod. Phys.}, 82:1287--1299, Apr 2010.

\bibitem{milian2015solitons}
C.~Mili{\'a}n, A.~V. Gorbach, M.~Taki, A.~V. Yulin, and D.~V. Skryabin.
\newblock Solitons and frequency combs in silica microring resonators:
  Interplay of the {R}aman and higher-order dispersion effects.
\newblock {\em Phys. Rev. A}, 92(3):033851, 2015.

\bibitem{PhysRevA.87.053852}
Y.~K. Chembo and C.~R. Menyuk.
\newblock Spatiotemporal {L}ugiato-{L}efever formalism for {K}err-comb
  generation in whispering-gallery-mode resonators.
\newblock {\em Phys. Rev. A}, 87:053852, May 2013.

\bibitem{Santhanam2003413}
J.~Santhanam and G.~P. Agrawal.
\newblock {R}aman-induced spectral shifts in optical fibers: general theory
  based on the moment method.
\newblock {\em Opt. Commun.}, 222(1–6):413 -- 420, 2003.

\bibitem{Zakharov&Shabat}
V.~E. Zakharov and A.~B. Shabat.
\newblock Exact theory of two-dimensional self-focusing and one-dimensional
  self-modulation of waves in nonlinear media (differential equation solution
  for plane self focusing and one dimensional self modulation of waves
  interacting in nonlinear media).
\newblock {\em Sov. Phys}, 34:62--69, 1972.

\bibitem{Barashenkov&Smirnov}
I.~V. Barashenkov and Y.~S. Smirnov.
\newblock Existence and stability chart for the ac-driven, damped nonlinear
  schrodinger solitons.
\newblock {\em Physical Review E}, 54:5707--5725, 1996.

\bibitem{PhysRevE.73.036621}
E.~N. Tsoy, A.~Ankiewicz, and N.~Akhmediev.
\newblock Dynamical models for dissipative localized waves of the complex
  {G}inzburg-{L}andau equation.
\newblock {\em Phys. Rev. E}, 73:036621, Mar 2006.

\bibitem{herr2015dissipative}
T.~Herr, M.~L. Gorodetsky, and T.~J. Kippenberg.
\newblock Dissipative {K}err solitons in optical microresonators.
\newblock In Philippe Grelu, editor, {\em Nonlinear Optical Cavity Dynamics:
  From Microresonators to Fiber Lasers}, chapter~6, pages 129--162. {John Wiley
  \& Sons}, 2015.

\bibitem{Nozaki1984}
K.~Nozaki and N.~Bekki.
\newblock Solitons as attractors of a forced dissipative nonlinear schrödinger
  equation.
\newblock {\em Physics Letters A}, 102(9):383 -- 386, 1984.

\bibitem{Wabnitz1993}
S.~Wabnitz.
\newblock {Suppression of interactions in a phase-locked soliton optical
  memory}.
\newblock {\em Opt. Lett.}, 18(8):601--603, Apr 1993.

\bibitem{Afanasjev:96}
V.~V. Afanasjev, C.~R. Menyuk, and Yu.~S. Kivshar.
\newblock Effect of third-order dispersion on dark solitons.
\newblock {\em Opt. Lett.}, 21(24):1975--1977, Dec 1996.

\end{thebibliography}

\end{document}